\documentclass[aps,pre,showpacs,twocolumn]{revtex4-1}

\usepackage{amsfonts}
\usepackage[dvips]{graphicx}

\begin{document}

\title{Analytical computation of the magnetization probability density function for the harmonic 2D XY model}
\author{G. Palma}
\email{guillermo.palma@usach.cl}
\author{D. Zambrano}
\affiliation{Departamento de F\'{\i}sica, Universidad de Santiago de Chile,\\
Casilla 307, Santiago 2, Chile.}

\begin{abstract}

The probability density function (PDF) of some global average
quantity plays a fundamental role in critical and highly correlated
systems. We explicitly compute this quantity as a function of the
magnetization for the two dimensional XY model in its harmonic
approximation. Numerical simulations and perturbative results have
shown a Gumbel-like shape of the PDF, in spite of the fact that the
average magnetization is not an extreme variable. Our analytical
result allows to test both perturbative analytical expansions and
also numerical computations performed previously. Perfect agreement
is found for the first moments of the PDF. Also for large volume and
in the high temperature limit the distribution becomes Gaussian, as
it should be. In the low temperature regime its numerical evaluation
is compatible with a Gumbel distribution.

\end{abstract}


\maketitle

The probability density function (PDF) of some global average
quantity plays a fundamental role in critical and highly correlated
systems. For example, it has been used in the so-called hypothesis
of universal fluctuations \cite{BHP}, which states that the PDF for
critical systems -like the two dimensional XY model- as a function
of the centered parameter of order divided by its standard deviation
should be a universal function $F$ of a Gumbel-type. This claim
should hold independent of the size and temperature for the magnetic
systems and Reynold number for turbulent experimental systems.
Approximated Gumbel distributions appear in quite dissimilar
physical systems, such as a confined turbulent flow or the roughness
$1/f$ noise in a resistor \cite{Racz}.

There was some evidence against this hypothesis reported in
\cite{COMM}. In particular, in Ref. \cite{MPV} for instance, an
analytical expression for the PDF for the full 2D XY model was
computed systematically by means of the loop expansion, and the
validity of the "generalized universality" has been linked to
renormalization group (RG) properties. The 2-loops analytical
expression for the PDF shows an explicit temperature dependence. As
a consequence, its skewness and kurtosis computed perturbatively up
to two loops (first order in $T$) show an explicit temperature
dependence. More recently \cite{P}, the third and fourth normalized
moments of the PDF -the skewness and kurtosis respectively- were
computed analytically in the two-dimensional harmonic XY model.
Their explicit temperature dependence was explicitly demonstrated,
which holds even in the thermodynamic limit. This result was an
indirect analytical proof of the failure of the claim of universal
fluctuations, allowing therefore to confirm the explicit temperature
dependence of the PDF reported in \cite{MPV}.

In spite of this, further papers supporting the proposed
"generalized universality" were published \cite{PHP}, raising the
question whether there is a underlying mechanism responsible for
this approximated phenomenon.

In a recent paper \cite{Bramwell} the shape of the different
distributions appearing in several critical systems has been
phenomenologically linked to scaling arguments and to the concept
from Renormalization Group Theory of classification of scaling
variables as irrelevant, marginal and relevant. Nevertheless, one
should note at this point that the so-called "generalized
universality" is not related to the concept of exact universality of
RG, because it only holds approximately, as it was explicitly shown
in ref. \cite{MPV} by using the loop -or temperature- expansion in
the context of the two-dimensional XY-model.

It is further stated in \cite{Bramwell} that an explicit analytical
computation of the PDF for the two dimensional XY model is still
missing. This claim also motivates our present computation.

Now we will explain how to compute a general analytical expression
for the PDF of the magnetization by using the spin wave or harmonic
approximation of the 2D XY model. This expression is valid for
arbitrary system size $L$ and temperature $T$. The 2D XY model
consists of planar spins $\phi _{\mathbf{x}}$ defined on a periodic
two dimensional square lattice $\Lambda $ of $N=L^2$ lattice sites,
which are coupled with nearest neighbors by cosine interactions.
According to RG arguments \cite{JKKN}, in the low temperature phase
and sufficiently below the Berezinskii-Kosterlitz-Thouless\ critical
temperature, the physics of this model is entirely described by its
harmonic approximation, the 2D HXY model. Indeed, in \cite{MPV} it
has been explicitly shown -by means of the loop expansion- that the
effect of the anharmonic corrections to the spin wave approximation
on the PDF is merely a renormalization of the temperature.
Nevertheless, and in spite of this fact, one should take into
account the periodicity of the variables of the XY model for the
boundary conditions, which leads to the contributions coming from
winding configurations \cite{MH}. This contribution turns out to be
numerically very small and therefore one expects to obtain with the
present model a trustable numerical approximation of the 2D XY model
in the large volume limit and in the low temperature phase.

The Hamiltonian of the 2D HXY-model is up to a constant
\begin{equation}
H(\phi )=\frac{1}{2}J\left\langle \phi ,-\Delta \phi \right\rangle
\end{equation}
where $\Delta$ is the Laplace operator on the lattice, $J$ is the
ferromagnetic constant and $\left\langle \phi ,\varphi \right\rangle
$ $ =\sum \phi (\mathbf{x})^{\ast }\varphi (\mathbf{x})$ stands for
the scalar product on the lattice. We use a system of units where
Boltzmann's constant is set equal to unity throughout the
computations and identify $T$ with the reduced temperature $T/J$.
Although this model has no phase transition, it is a critical model
in the sense that it has an infinite correlation length. This
Gaussian model has been analytical extensively studied, because it
represents the starting point for perturbative expansions and due to
the involvement of Gaussian integrals. Some useful physical
quantities can be expressed in terms of the Fourier representation
of the lattice propagator $G$
\begin{equation}
G(\mathbf{x})=\frac{1}{N}\sum\limits_{(\mathbf{K}_{L}\ )^{2}\neq
0}\frac{\exp (-i\mathbf{K}\cdot \mathbf{x})}{(\mathbf{K}_{L})^{2}}\
\label{GLat}
\end{equation}
where $\mathbf{K}_{L}$ is the lattice momentum defined as usual as
$(K_{L})_{i}\ =2\sin (K_{i}/2),$ with $i=1,2$ and $K_{i}$ lies in
the first Brillouin zone, $K_{i}=(2\pi /L)n$ with $n\in \mathbb{Z}$
and $-\pi <K_{i}\leq \pi $. The sum runs over all possible values of
$K_{i}$ for which ($\mathbf{K}_{L})^{2}$ does not vanish. This comes
from the fact that the Goldstone mode, which is originated by the
invariance of the original Hamiltonian under a global rotation of
the spin variables, and which leads to the translation invariance on
the lattice, must be removed from the calculation. As it was first
shown in Ref. \cite{MPV}, the PDF can be defined as the Fourier
transform of the partition function $Z(q)$ of an auxiliary theory,
which differs from the original theory by a dimension 0 perturbation
with a very small imaginary coefficients $iq/N$. This theory turns
out to be asymptotically free in the infrared (this is the case of
the 2D XY model), viz.
\begin{equation}
P(M)=\int_{-\infty }^{\infty }\frac{dq}{2\pi }\exp
\{iq(M-\left\langle M\right\rangle \}Z(q)  \label{PDF}
\end{equation}
where
\begin{equation}
Z(q)=\frac{ e^{-iq\left\langle M\right\rangle } }{Z_{0}}\int D\phi
\exp \bigg\{-\beta H(\phi )+\frac{iq}{N}\sum\limits_{x\in \Lambda
}\cos \phi _{x} \bigg\}  \label{Zq}
\end{equation}
with partition function $Z_{0}=\int D\phi \exp \{-\beta H(\phi )\}$.
The mean $\left\langle M\right\rangle $ and the higher order moments
of the PDF are obtained as usual as the integrals $\left\langle
M^{p}\right\rangle =\int M^{p}P(M)dM$. In particular, the mean
square fluctuation is defined by
$\sigma ^{2}=\left\langle (M-\left\langle M\right\rangle )^{2}\right\rangle $%
. The partition function $Z(q)$ can be written as an integral over
the
normalized Gaussian measure $d\mu _{TG}(\phi )$ of covariance $TG$:%
\begin{equation}
Z(q)=\exp (-iq\left\langle M\right\rangle )\int d\mu _{TG}(\phi )\exp \bigg\{%
\frac{iq}{N}\sum\limits_{x\in \Lambda }\cos \phi _{x}\bigg\},
\label{ZqTG}
\end{equation}
where $G$ denotes the lattice propagator given by eqn.(\ref{GLat}).

In terms of the PDF, the hypothesis of BHP universality stipulates
that the PDF considered as a function of the reduced variable $\mu
=(M-\left\langle M\right\rangle )/\sigma $ is a universal function
$F$, and should be the same for a wide class of strongly correlated
critical systems, independent of the system temperature $T$ and
volume $N$, provided $N$ is large enough:
\begin{equation}
F(\mu )=\sigma P\bigg\{\frac{(M-\left\langle M\right\rangle
)}{\sigma }\bigg\}. \label{Funiv}
\end{equation}

\begin{figure}[ht]
\centering
\includegraphics[scale=0.5]{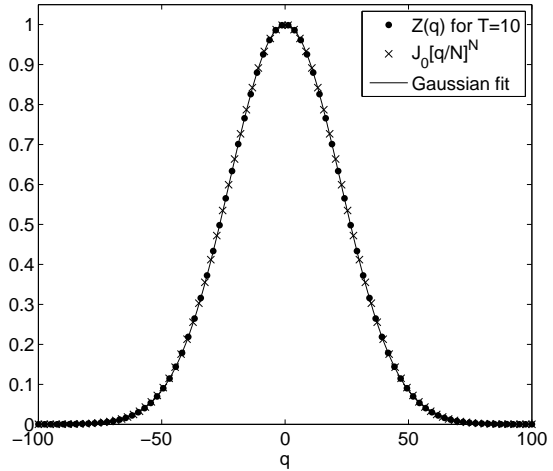} \caption{In this figure we plot the
approximate expression for $Z(q)$ given by equation
(\ref{Zqfinal_hT}) for $L = 16$. $Z(q)$ is also evaluated
numerically from eqn.(\ref{Zqfinal}) for a lattice with $L = 16$ and
$T = 10$. In order to compare both results and to obtain an
analytical simple expression a Gaussian fit is performed obtaining
$\chi^{2} = 1$.} \label{fig.1}
\end{figure}

Moreover, this universal function should be non-Gaussian and has a
functional dependence very close to the one corresponding to a
Gumbel distribution for certain exponent $a$, which is known in the
context of extremum statistics \cite{CRW}. We notice that up to the
first factor, the partition function $Z(q)$ defined by the integral
(\ref{Zq}) corresponds to the partition function of the Sine-Gordon
theory with a small imaginary fugacity $z=iq/N$. It is
straightforward to see that $Z(q)$ is the generating functional of
the centered high order moments of $P$, \textit{i.e.}
\begin{equation}
i^{n}\left[ \frac{d^{n}}{dq^{n}}Z(q)\right] _{q=0}=\left\langle
(M-\left\langle M\right\rangle )^{n}\right\rangle . \label{Mp}
\end{equation}
Motivated by the method used to study the renormalization group flow
in the Sine-Gordon model \cite{JW}, we will use the Fourier
representation of an exponential factor (which is also known from
the high temperature -or cluster- expansion of the XY model
\cite{ID}),
\begin{equation}
\exp \{i\lambda \cos \phi \}=\sum\limits_{m=-\infty }^{\infty
}i^{m}J_{m}(\lambda )\exp \{im\phi \},  \label{Fourier}
\end{equation}
where $J_{m}(\lambda )$ are the Bessel function of integer order
$m$, and \ the functional identity for the generating function of
the Gaussian measure $d\mu _{TG}(\phi )$ \cite{GJ}:
\begin{equation}
\int d\mu _{TG}(\phi )\exp (i\left\langle \phi ,f\right\rangle
)=\exp (-\frac{1}{2}\left\langle f,TGf\right\rangle ),
\end{equation}
to compute analytically the integral of eqn. (\ref{ZqTG}). The
resulting expression for $Z(q)$, which is our main analytical
result, reads
\begin{eqnarray}
Z(q) &=&e^{-iq\left\langle M\right\rangle
}\sum\limits_{m_{1},m_{2,}..,m_{N}}\prod\limits_{k=1}^{N}\left(
i^{m_{k}}J_{m_{k}}(q/N)\left\langle M\right\rangle
^{m_{k}^{2}}\right)
\nonumber \\
&&\times \exp \left\{
-T\sum\limits_{i<j}m_{i}m_{j}G(\textbf{x}_{i}-\textbf{x}_{j})\right\}
\label{Zqfinal}
\end{eqnarray}
From this explicit expression for $Z(q)$ and using eqn. (\ref{Mp})
one can compute explicitly the moments of the PDF, obtaining for
instance
\begin{equation}
\left\langle M\right\rangle =\exp \bigg\{-\frac{TG(0)}{2}\bigg\}
\label{M}
\end{equation}
\begin{equation}
\sigma =\left\langle M\right\rangle \
\bigg\{\frac{1}{N}\sum\limits_{\textbf{z}\in \Lambda }(\cosh
[TG(\textbf{z})]-1)\bigg\}^{1/2} \label{Sigma}
\end{equation}
\begin{eqnarray}
\left\langle M^{3}\right\rangle  &=&\frac{\left\langle M\right\rangle ^{3}}{%
2N^{2}}\sum_{\mathbf{x},\mathbf{y}\in \Lambda }  \label{M3} \\
&&\times {\left( e^{-TG(\mathbf{x})}\cosh {\left\{ T\left[ G(\mathbf{y})+G(%
\mathbf{x}-\mathbf{y})\right] \right\} }\right. }  \nonumber \\
&&\ \ \ \ \left. +e^{TG(\mathbf{x})}\cosh {\left\{ T\left[ G(\mathbf{y})-G(%
\mathbf{x}-\mathbf{y})\right] \right\} }\right) .  \nonumber
\end{eqnarray}
These expressions are exact and agree with previous results reported
in \cite{P}. Higher centered moments of the PDF can be computed as
well by using equations (\ref{Mp}) and (\ref{Zqfinal}). The result
perfectly agrees with their corresponding expressions reported for
instance in Ref. \cite{P} (see eqns. (2.4) and (2.5)).

In order to obtain the analytical expression for the "universal
distribution function $F$" one has to insert the expression for
$Z(q)$ given by eqn. (\ref{Zqfinal}) into (\ref{PDF}), which leads
to the result

\begin{equation}
F(\mu ;T)=\int \frac{dQ}{2\pi }\exp \{iQ\mu \}Z(-Q/\sigma ) ,
\label{Fmu}
\end{equation}
where $\left\langle M\right\rangle $ and $\sigma $ are given by
eqns. (\ref{M}) and (\ref{Sigma}).

Now we want to study numerically the expression for the PDF deduced
for the 2D HXY model as a function of the system temperature $T$ and
volume $N$. From the equation (\ref{M}) it follows that in the high
temperature limit, the mean magnetization goes to zero. In this
limit, the exponential factor of equation (\ref{Zqfinal}) vanishes
for all values of $m_k\in Z$ except for $m_k=0$. Therefore, in this
limit equation (\ref{Zqfinal}) goes into

\begin{equation}
Z(q)\big|_{T \rightarrow \infty} = \left[J_{0}(q/N)\right]^{N}
\label{Zqfinal_hT}
\end{equation}

Moreover, for large volume the above expression for $Z(q)$ becomes
Gaussian. This point can be shown analytically performing a Taylor
expansion valid for small arguments of $J_{0}(q/N)$. Within this
approximation, the mean square fluctuation of $Z(q)$ is given by
$\sigma\approx\sqrt{2N}$. On the other side, $Z(q)$ can be also
evaluated numerically by performing the sums appearing in
eqn.(\ref{Zqfinal}) over the relevant configurations of
$m_k$-values. Both curves are displayed in Fig.\ref{fig.1}, for a
square lattice of lattice size $L=16$ and $T=10$. They felt onto the
same curve with remarkable accuracy. Perfect agreement is found when
a Gaussian distribution is fitted to $Z(q)$, with a value of
$\chi^{2} = 1$. Also the numerical value for $\sigma = 22.51$ agrees
with the analytical expression for $\sigma$ in the high-T limit.

\begin{figure}[ht]
\centering
\includegraphics[scale=0.5]{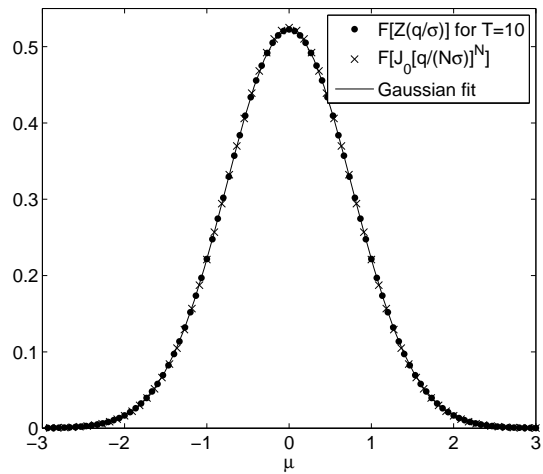} \caption{This figure shows the
inverse Fourier transform of $Z(q)$ -or equivalently the PDF defined
by eq. (\ref{Fmu})- corresponding to the data appearing in figure
\ref{fig.1}.} \label{fig.2}
\end{figure}

\begin{figure}[ht]
\centering
\includegraphics[scale=0.5]{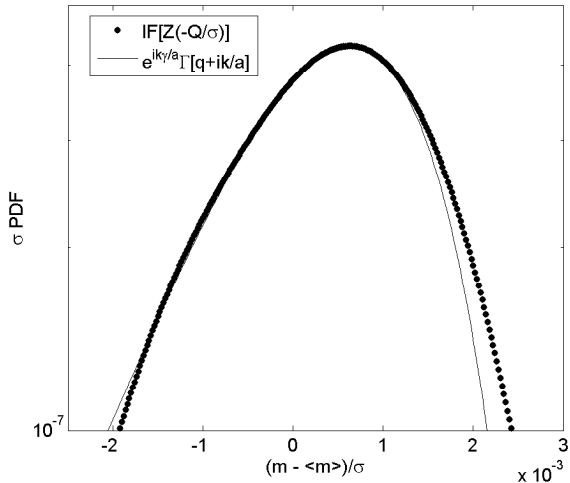} \caption{This figure shows with a
dashed line the numerical inverse fourier transform of $Z(q)$ given
by eq. (\ref{Fmu}) for lattice size $L = 16$ and temperature $T =
0.7$. For comparison we plot (full line) the analytical expression
reported in ref. \cite{Racz}, found for the roughness $1/f$ noise.
The agreement between both curves is noteworthy.} \label{fig.3}
\end{figure}

The PDF itself can be computed both numerically and analytically in
the high temperature limit. Indeed, using the approximated numerical
values obtained already for $Z(q)$, we evaluate its Fourier
transformation defined by the integral of Eqn.(\ref{PDF}). One may
use instead the accurate analytical Gaussian expression for $Z(q)$
found in this limit to perform analytically the integral of
Eqn.(\ref{PDF}). Both results are in perfect agreement as it is
shown in figure \ref{fig.2}, where a lattice of lattice size $L=16$
and temperature $T=10$ was used. The values obtained for the
skewness and kurtosis, $s\approx 0$ and $c\approx 3$ respectively,
perfectly agree, within only a few percent of error with their
corresponding values reported in Ref. \cite{P}.

Finally, we can obtain the PDF itself in the low temperature regime
using direct numerical integration. In figure 3 a numerical
evaluation of the distribution for the reduced magnetization is
shown  for a lattice of lattice size $L = 16$ and temperature $T =
0.7$ (dashed line). This corresponds to the inverse fourier
transform of $Z(q)$ given by eq. (\ref{Fmu}). In order to compare
our expression with the analytical expression for the roughness of
signals displaying $1/f$ power spectra reported in ref. \cite{Racz},
we also plot (full line) the corresponding Gumbel-like distribution.
The agreement between both curves is noteworthy.

\acknowledgments Partial support by DICYT, University of Santiago
grant 040931PA, is gratefully acknowledged.


\begin{thebibliography}{0}

\bibitem{BHP} S. T Bramwell, P. C.W. Holdsworth and J. F. Pinton, Nature (London) \textbf{396} (1998) 552.

\bibitem{Racz} T. Antal, M. Droz, G. Gy\"{o}rgyi \and Z. R\'{a}cz, Phys. Rev. Lett. \textbf{87} (2001) 240601.

\bibitem{COMM} B. Zheng and S. Trimper, Phys. Rev. Lett. \textbf{87}, 188901
(2001); V. Aji and N. Goldenfeld, Phys. Rev. Lett. \textbf{86}, 1007
(2001);
T. Antal, M. Droz, G. Gy\"{o}rgyi, and Z. R\'{a}cz, Phys. Rev. Lett. \textbf{%
87}, 240601 (2001); N. W. Watkins, S. C. Chapman and G. Rowlands,
Phys. Rev. Lett. \textbf{89}, 208901 (2002); G. Palma, T. Meyer and
R. Labb\'{e}, cond-mat/0007289 (2000), and Phys. Rev. E \textbf{66},
026108 (2002).

\bibitem{MPV} G. Mack, G. Palma and L. Vergara, Phys. Rev. E \textbf{72},
026119 (2005).

\bibitem{P} G. Palma, Phys. Rev. E \textbf{73}, 046130 (2006).

\bibitem{PHP} P. C. W. Holdsworth and M. Sellitto, Physica A \textbf{315},
643 (2002); B Portelli et al., J. Phys. A: Math. Gen. \textbf{35},
1231 (2002); S.T. Bramwell, T. Fennel, P.C.W. Holdsworth and B.
Portelli, Europhys. Lett. \textbf{57}, 310 (2002); B. Portelli, P.
C. W. Holdsworth, and J.-F. Pinton, Phys. Rev. Lett. \textbf{90},
104501 (2003); M. Clusel, J.-Y. Fortin, and P. C. W. Holdsworth,
Phys. Rev. E \textbf{70}, 046112 (2004).

\bibitem{Bramwell} S. T. Bramwell, Nature (London) \textbf{5} (2009) 443.

\bibitem{JKKN} J. V. Jos\'{e}, L. P. Kadanoff, S. Kirkpatrik and D. R.
Nelson, Phys. Rev. B \textbf{16}, 1217 (1977).

\bibitem{MH} M. Hasenbush, J. Phys. A \textbf{38 }(2005) 5869.

\bibitem{CRW} S. C. Chapman and G. Rowlands and N. W. Watkins, J. Phys. A:
Math. Gen. \textbf{38}, 2289 (2005).

\bibitem{JW} J. Wuerthner, Diplomarbeit am II. Institut fuer Theoretische
Physik der Universitaet Hamburg, Hamburg (1997).

\bibitem{ID} C. Itzykson and J.-M. Drouffe, Statistical field theory,
(Cambridge University Press 1989).

\bibitem{GJ} J. Glimm and A. Jaffe, Quantum Physics: A Functional Integral
Point of View (Springer, New York, 1987).

\end{thebibliography}
\end{document}